\documentclass[conference]{IEEEtran}
\IEEEoverridecommandlockouts
\usepackage{cite}
\usepackage{amsmath,amssymb,amsfonts}
\usepackage{algorithmicx,algpseudocode}
\usepackage{program}
\usepackage{algorithm}
\usepackage{graphicx}
\usepackage{textcomp}
\usepackage{xcolor}
\usepackage{multirow}
\def\BibTeX{{\rm B\kern-.05em{\sc i\kern-.025em b}\kern-.08em
    T\kern-.1667em\lower.7ex\hbox{E}\kern-.125emX}}
\begin{document}

\title{F-RouND: Fog-based Rogue Nodes Detection in Vehicular Ad hoc Networks\\}


\author{\IEEEauthorblockN{Anirudh Paranjothi\IEEEauthorrefmark{1},
Mohammed Atiquzzaman\IEEEauthorrefmark{1}, Mohammad S. Khan \IEEEauthorrefmark{2}}
\IEEEauthorblockA{School of Computer Science, University of Oklahoma, Norman, Oklahoma, USA\\
\IEEEauthorblockA{Department of Computing, East Tennessee State University, Johnson City, Tennessee, USA\\
e-mail: \IEEEauthorrefmark{1}\{anirudh.paranjothi, atiq\}@ou.edu, \IEEEauthorrefmark{2}khanms@etsu.edu}}}

\maketitle

\begin{abstract}
Vehicular ad hoc networks (VANETs) facilitate vehicles to broadcast beacon messages to ensure road safety. The rogue nodes in VANETs broadcast malicious information leading to potential hazards, including the collision of vehicles. Previous researchers used either cryptography, trust values, or past vehicle data to detect rogue nodes, but they suffer from high processing delay, overhead, and false-positive rate (FPR). We propose fog-based rogue nodes detection (F-RouND), a fog computing scheme, which dynamically creates a fog utilizing the on-board units (OBUs) of all vehicles in the region for rogue nodes detection. The novelty of F-RouND lies in providing low processing delays and FPR at high vehicle densities. The performance of our F-RouND framework was carried out with simulations using OMNET++ and SUMO simulators. Results show that F-RouND ensures 45\% lower processing delays, 12\% lower overhead, and 36\% lower FPR at high vehicle densities compared to existing rogue nodes detection schemes.
\end{abstract}

\,


\section{Introduction}

Vehicular ad hoc networks (VANETs) are considered as the keystone of intelligent transportation systems (ITS) to enhance road safety by reducing the number of accidents and optimizing the traffic flow. VANETs provide vehicle-to-vehicle (V2V), and vehicle-to-infrastructure (V2I) communication depends on dedicated short range communication (DSRC), which consists of a set of protocols for transmitting the messages between vehicles and between vehicles and the roadside infrastructures, such as roadside units (RSUs), etc. [1, 2]. The vehicles are equipped with on-board units (OBUs) for transmitting and receiving the messages, including beacon messages. Beacon messages are broadcasted in VANETs to disseminate network state or emergency information, which can be utilized to reduce road accidents and traffic congestion [3]. However, rogue vehicles, also known as rogue nodes broadcast malicious information, such as false congestion information and collision warning by broadcasting low-speed values in beacon messages to change the normal behavior of the vehicles may lead to catastrophic consequences, such as the collision of vehicles [4].  Detecting rogue nodes plays a crucial role in establishing a secure VANETs environment.

Previous authors used either cryptography, trust values, or past vehicle data to detect rogue nodes. Al-Otaibi et al. [5] presented a cryptography-based intrusion detection scheme (IDS) using fog computing. The proposed scheme considers RSUs as fog nodes for rogue nodes detection. However, the approach [5] encounters a high processing delay and overhead in detecting rogue nodes when the RSUs are overloaded or not available in the region. Zaidi et al. [6] proposed an IDS to detect rogue nodes based on past vehicle data. Each vehicle utilizes its OBU to detect false data propagated by the rogue nodes. Ahmad et al. [7] proposed a trust-based scheme termed as trust evaluation and management (TEAM). The TEAM framework consists of three different trust models: entity-oriented, data-oriented, and hybrid-oriented trust models for detecting rogue nodes.  RSUs are used to compute trust scores and segregates the vehicles based on the calculated trust score. The existing approaches [6, 7] have limitations, such as high delay, overhead, and false-positive rate (FPR).

To address the shortcomings of the existing rogue nodes detection schemes, we introduce an OBU-based dynamic fog computing technique called fog-based rogue nodes detection (F-RouND). The F-RouND framework employs a two-fold process in rogue nodes detection: first, we use the concept of guard node to detect rogue nodes. The guard node is the vehicle that has more neighboring vehicles in its transmission range, dynamically creates a fog utilizing the OBUs of all vehicles in the region, and then the dynamic fog is used to compare the received beacon messages from all vehicles to detect rogue nodes. Second, the guard node performs the hypothesis test to validate whether the rogue nodes are correctly identified or not. Upon successful validation, the guard node broadcasts the information of rogue nodes to all vehicles in the region. We adopt fog computing, as it offers unique services, including low latency and high bandwidth compared to traditional communication techniques [8].

The \textit{difference} between the F-RouND framework and existing schemes is, each vehicle uses either its OBU or RSU to detect rogue nodes [5-7]. RSUs are deployed only in the critical regions of the road. The absence of RSUs yields high processing delay and FPR. OBUs of an individual vehicle is highly resource-constrained encounters a high delay in analyzing the data at high vehicle densities. Whereas, in the F-RouND framework, the guard node combines OBUs of all vehicles in the region in creating the dynamic fog. Utilizing OBUs of all vehicles increases the computational power of dynamic fog resulting in low processing delay and FPR.

Our \textit{objective} is to reduce latency, increase true positive rate (TPR), and decrease FPR in detecting rogue nodes at high vehicle densities. We considered three existing rogue nodes detection schemes for comparison:  Fog-IDS [5], IDS [6], and TEAM [7]. The performance of our framework was carried out using OMNET++ and SUMO simulators. Our results lead to an exciting conclusion that the F-RouND framework reduces the latency and FPR, and performs up to 38\% better than the existing rogue nodes detection schemes [5-7].

The \textit{contributions} of the paper are: 1) We proposed a framework that uses statistical techniques and traffic models to detect rogue nodes in VANETs with low delay, low FPR, and high TPR. 2) We introduced the guard node in the F-RouND framework, which uses an OBU-based fog computing technique to compare and validate the received beacon messages from all vehicles in the region. 3) The proposed framework does not depend on any roadside infrastructures, including RSUs in rogue nodes detection.

The rest of the paper is structured as follows: related work is discussed in Section II. The proposed solution for rogue nodes detection is presented in Section III. Section IV evaluates the performance of our approach through extensive simulation. Finally, conclusions and future work are given in Section V.

\section{Related Work}

This section presents an overview of the most recent existing schemes that detect rogue nodes in VANETs. Arshad et al. [9] proposed a beacon-based trust scheme to detect false messages in VANETs. Initially, the trust values of all vehicles are assigned to be 0, and then based on the correctness of the data, positive or negative trusts are assigned. The calculated trust of any vehicle reaches a predefined threshold limit are known as rogue nodes, and then the information is broadcasted to all the vehicles in the region. However, the proposed work [9], suffers from high packet loss ratio (PLR) and FPR.

Sedelmaci et al. and Ahmed et al. [10, 11] proposed trust-based schemes to detect rogue nodes. RSUs are used to compute trust scores and detect rogue nodes based on the calculated trust score. Yang et al. [12] proposed a tree-based machine learning algorithm to classify whether received data is valid or not based on the historical vehicle data. The results of the classification are then combined to detect rogue nodes broadcasted false information. The frameworks [10-12] encounter a low TPR and high FPR in detecting rogue nodes. 

Zhang et al. and Shams et al. [13, 14] illustrated the rogue nodes detection mechanism based on the support vector machine (SVM)  to resist false messages. The proposed mechanism [13] consists of a local trust module and a vehicle trust module, where the local trust module uses an SVM-based classifier to detect false messages, and the vehicle trust module uses SVM to derive comprehensive trust value for all vehicles. Finally, the results of both the local module and the trust module are then combined to find rogue nodes in the region. The authors [14] use the SVM-based module to analyze the past vehicle data. Based on the analysis, the trust values of the vehicle are calculated, which in turn used for detecting the rogue nodes. However, these approaches [13, 14] have high delay and overhead at high vehicle densities.

To overcome the limitations of the existing rogue nodes detection schemes [5-7, 9-14], we propose the F-RouND framework, which uses fog computing technique to detect rogue nodes in VANETs. The proposed framework does not depend on either trust score or past vehicle data and performs better even when there are 40\% rogue nodes in the region.

\begin{figure}[tbp]
\centering
\includegraphics[width=250pt]{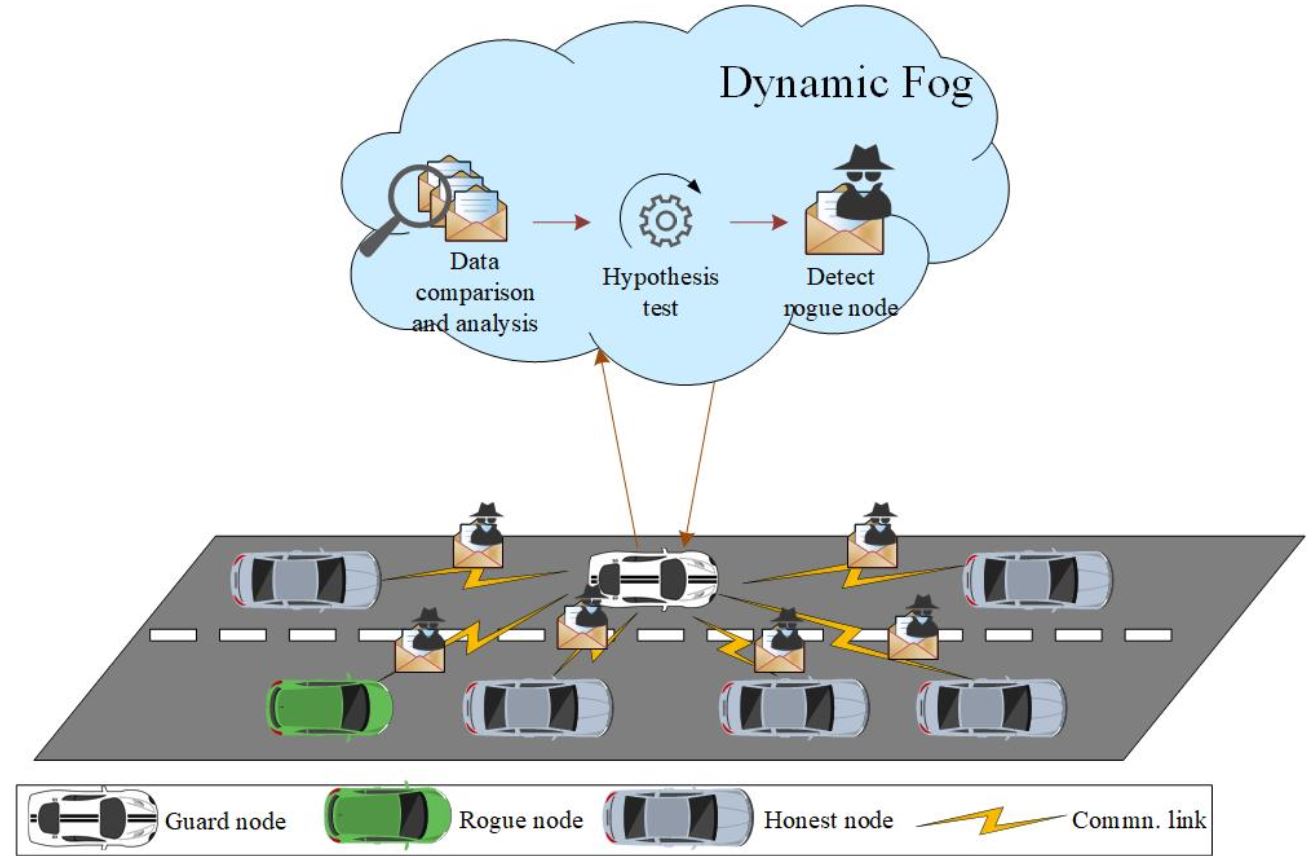}
\caption{Execution scenario of the F-RouND framework in the presence of a rogue node using fog computing technique.\label{fig1}}
\end{figure}

\section{Proposed F-RouND Framework}
In this section, we provide the working principle of our proposed F-RouND framework. The F-RouND framework engages fog computing technique to detect rogue nodes due to the fog can be created at the proximity of users and performs computations at the edge of the network [8]. In order to detect rogue nodes, F-RouND employs the concept of guard node. 

The vehicle which has a greater number of the neighboring vehicle in its transmission range will act as a guard node. Guard node dynamically creates a fog using OBUs of all vehicles to compare and analyze vehicle speeds to detect rogue nodes. All the vehicles in the region broadcast similar speed values as they are in similar traffic conditions and dependent on other vehicles under all circumstances. Thus, if there is a significant difference in vehicle speed, the guard node considers the vehicles as rogue nodes, and then the hypothesis test is carried out to validate whether the rogue nodes are correctly identified or not. If the hypothesis test yields speed values within the acceptance range, then the vehicles are considered as honest nodes. Otherwise, the vehicles are highlighted as rogue nodes. The guard node broadcasts the information of rogue nodes to all vehicles in the region to ignore the beacon messages received further from the rogue nodes. One such scenario of our F-RouND framework is depicted in Fig. 1. The computation power of the guard node increases when the number of vehicles increases as the OBUs of all vehicles are utilized in creating a dynamic fog results in lower delay compared to [5-7] rogue node detection schemes.

\subsection {Selection of Guard Node}

Rogue nodes are the vehicles broadcasting low-speed values to change the normal behavior of the vehicles for own benefits. The guard node analyzes received beacon messages from all vehicles to detect rogue nodes. The following three assumptions are made in the selection of the guard node: first, we assume that the center vehicle has a greater number of the neighboring vehicles in its transmission range compared to the front and tail-end vehicles. Hence, we select the center vehicle as a guard node. Second, we assume that the guard node is the most trustworthy vehicle in the network. Thus, the guard node cannot be turned out to be a rogue node under any circumstances. Third, we assume that the total number of vehicles (\textit N) in the region at any given time is at least two as the guard node needs at least two vehicle data to compare and analyze the beacon messages to detect rogue nodes. 

The guard node of the F-RouND framework is selected as follows: initially, we take the mean of position vectors of all vehicles (i.e., $P_1$, $P_2$, ...., $P_N$) to find a unique center point $\zeta$, and then we calculate Euclidean distance between $\zeta$ and the position vector of each vehicle to determine the point that has the minimum distance from $\zeta$. Finally, the vehicle located at this point will be selected as the guard node, $G_{veh}$.

\begin{equation} \label{eq11}
 \begin{split}
 \zeta = \frac{1}{N} \sum_{i=1}^{N} P_i
 \end{split}
 \end{equation}

\begin{equation} \label{eq12}
 \begin{split}
 G_{veh} = \arg\min_{P_i \in X} \|\zeta - P_i\|
 \end{split}
 \end{equation} 
 
  Where, $X$ = \{$P_1$, $P_2$....,$P_N$\}.

\subsection {Speed and Density of Vehicles}

The vehicles broadcast beacon messages every 100 ms. In the F-RouND framework, Greenshield's mathematical model is utilized to model the traffic flow in the region. Greenshield traffic model is considered to be a fairly accurate and simple model for real-world traffic flows works under the assumption of density ($\rho$), and the speed of the vehicles (\textit S) is negatively correlated [7]. The density can be calculated as:

\begin{equation} \label{eq3}
 \begin{split}
 \rho = B_{msg} \cdot N
 \end{split}
 \end{equation} 
 
 Where $B_{msg}$ is the beacon message broadcasted from one vehicle id and $N$ is the total number of vehicles in the region. As the speed and density of the vehicles are negatively correlated, the density increases when the speed of the vehicles decreases in the region. The relationship between speed and density can be defined as:

 \begin{equation} \label{eq5}
 \begin{split}
 S = S_{max} -  \frac {\rho}{\rho_{max}} S_{max}
 \end{split}
 \end{equation}

 Where $S_{max}$ is the speed of the vehicle when density is zero and ${\rho_{max}}$ is the maximum density, also point at which speed of the vehicles becomes zero. In addition to usual parameters, such as speed, acceleration, braking status, location, gap, VIN, etc., the beacon message of all vehicles in the F-RouND framework also includes the density information ($\rho$).

 The selected guard node (Section 3A) creates a dynamic fog to compare and analyze the vehicle speed in beacon messages to detect rogue nodes in the region. Once the rogue nodes are identified, the guard node calculates the average density ($\rho_{avg}$) and the average speed ($ S_{avg}$) as follows:
\begin{equation} \label{eq4}
 \begin{split}
 \rho_{avg} = \frac{1}{N} \sum_{i=1}^{N} \rho_{i}
 \end{split}
 \end{equation} 
 
  \begin{equation} \label{eq7}
 \begin{split}
 S_{avg} = \frac{1}{N} \sum_{i=1}^{N} S_{i}
 \end{split}
 \end{equation} 
 
The guard node uses the average speed ($ S_{avg}$) to perform the hypothesis test. During the hypothesis test, vehicles that correspond with average speed are termed as honest nodes. In case the average speed difference is high or low, the upper and lower bound values are calculated to decide whether a received speed value should be accepted or not. The hypothesis test provides a significant contribution in reducing FPR at high vehicle densities compared to [5-7] schemes.

\subsection {Hypothesis Test to Validate the Speed of Vehicles}

\begin{table}[tbp]
\centering
\caption{Types of Error and Decisions in Null Hypothesis Testing \label{tab1}}
\begin{tabular}{|l|l|l|l|}
\hline
\multicolumn{2}{|l|}{\multirow{2}{*}{}}                                                                                                  & \multicolumn{2}{c|}{\textbf{Null hypothesis ($H_0$)}}                                                                                                    \\ \cline{3-4} 
\multicolumn{2}{|l|}{}                                                                                                                   & \multicolumn{1}{c|}{\textbf{True}}                                      & \multicolumn{1}{c|}{\textbf{False}}                                       \\ \hline
\multicolumn{1}{|c|}{\multirow{2}{*}{\textbf{\begin{tabular}[c]{@{}c@{}}Null hypothesis \textbf{($H_0$)} \\     decision\end{tabular}}}} & \textbf{Accept} & No error                                                                & \begin{tabular}[c]{@{}l@{}}Type II error \\ (False negative)\end{tabular} \\ \cline{2-4} 
\multicolumn{1}{|c|}{}                                                                                                 & \textbf{Reject} & \begin{tabular}[c]{@{}l@{}}Type I error\\ (False positive)\end{tabular} & No error                                                                  \\ \hline
\end{tabular}
\end{table}

Hypothesis testing allows a confidence interval to the range of values that allows us to accept a claim with a certain confidence. The F-RouND framework performs a hypothesis test with the speeds received from all vehicles in the region, which allows the guard node to accept the speeds with a certain confidence. Moreover, hypothesis testing is a commonly used statistical technique when we have two different claims, of which only one claim can be true at any given time. In the F-RouND framework except for the guard node, we have two different claims for all the vehicles in the region, i.e., either the vehicle is honest or rogue. If the vehicle is honest, the guard node accepts the data, else if the vehicle is rogue, the guard node rejects the data, and then the information of the rogue nodes is broadcasted to all the vehicles in the region. We use the hypothesis test to validate if the vehicle is honest or rogue using speed values in the beacon messages.

There are two hypotheses involved in the hypothesis testing approach: the null hypothesis ($H_0$) and the alternate hypothesis ($H_a$). The null hypothesis is the claim that needs to be tested, and the alternate hypothesis is everything else. If the null hypothesis is accepted, then the alternate hypothesis is rejected, and vice versa. In the F-RouND framework, the null hypothesis ($H_0$) is that the speed value received is from an honest vehicle, and the alternate hypothesis  ($H_a$) is that the speed value received is from a rogue node. Two types of error associated with the hypothesis testing approach: the first type of error (Type I error) occurs when the null hypothesis is wrongly rejected, also known as a false positive, and the second type of error (Type II error) occurs when the null hypothesis is wrongly not rejected, also known as a false negative, as shown in Table I. False negative is not as severe as false positive as it may not lead to any catastrophic consequences in the network. 

\begin{figure}[tbp]
\includegraphics[width=250pt, height=11pc]{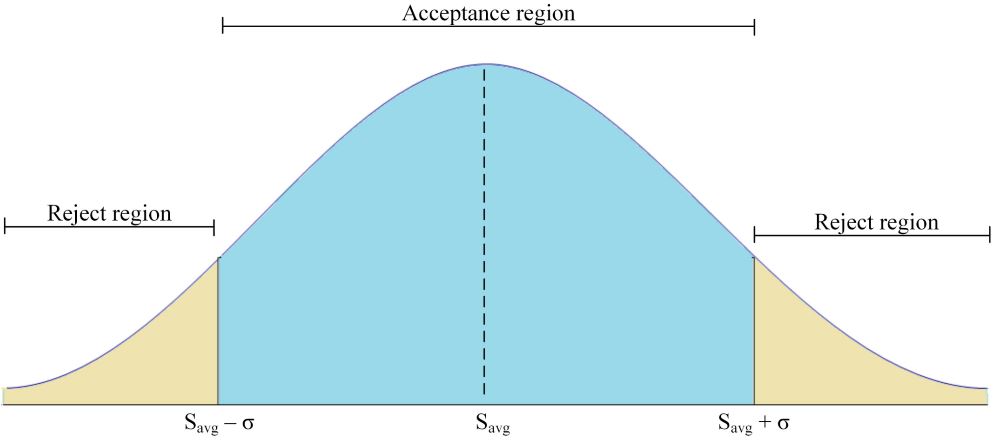}
\caption{Hypothesis test of F-RouND framework based on the average vehicle speed to determine acceptance range values.}
\end{figure}

We use standard deviation $(\sigma)$ to measure the variation of average speed with speed values received from all vehicles. 
 \begin{equation} \label{eq8}
 \begin{split}
 \sigma = \sqrt {\frac{1}{N} \sum_{i=1}^{N} {(S_{avg} - S_{i})}^2}
 \end{split}
 \end{equation} 

A low standard deviation indicates that the speed values received from the vehicles tend to be close to the average speed calculated by the guard node, while a high standard deviation indicates that the speed values received from the vehicles have highly deviated from the average speed. The upper and lower limits of our acceptance region will be $S_{avg} - \sigma$ and $S_{avg} + \sigma$. The speed values received from the honest nodes fall in the acceptance region  $S_{avg} - \sigma < S_{avg} < S_{avg} + \sigma$ when the null hypothesis is true, as shown in Fig. 2. Thus, the speed values received outside the acceptance region are rejected (i.e., $S_{avg} - \sigma > S_{avg} > S_{avg} + \sigma$).

Unlike existing rogue nodes detection schemes [5-7, 9-14], the F-RouND framework works efficiently for all vehicle densities as well as for all road conditions. For example, in case of an accident or high dense downtown regions, the speed of all vehicles drop, which will decrease the average speed ($ S_{avg}$) for the region, and as a result, the speed values of the honest nodes remain in the acceptance region. Once the rogue nodes are identified, the guard node includes the rogue node id and the result of the hypothesis test, i.e., either 0 or 1 (Eqn. 8) to the beacon messages, and then broadcast the beacon messages to all vehicles in the region. All vehicles in the region start ignoring the beacon messages received further from the rogue nodes to contain the damage. 
 \begin{equation}
  Rlt =
    \begin{cases}
      0; & \text{${S_{avg} - \sigma < S_{avg} < S_{avg} + \sigma}$}\\
      1; & \text{Otherwise}
    \end{cases}       
\end{equation}

\subsection {F-RouND Algorithm}
 
\begin{algorithm}
\caption{F-RouND - Rogue nodes detection algorithm }\label{alg1}
\textbf{Input:} $G_{veh}$ receives $B_{msg}$ from all vehicles in the region
\newline
\textbf{Output:} $G_{veh}$ broadcasts information of rogue nodes

\begin{algorithmic} [1]
\If {($N$ $\geq$ 2)} 
\State Calculate  $\zeta$ and Euclidean distance 
\State Assign  $G_{veh}$ 
\Else \, GoTo 19
\EndIf
\State $G_{veh}$ dynamically creates a fog 
\State $G_{veh}$ receives $B_{msg}$ from all vehicles in the region
\For {each  $B_{msg}$  received} 
\State Calculate $S_{avg}$ and  $\rho_{avg}$
\State Perform hypothesis test 
\If {$S$ in the acceptance range}
\State Declare the vehicle as honest node
\Else
\State Declare the vehcile as rogue node
\State Store the rogue node id
\EndIf
\EndFor
\State $G_{veh}$ broadcasts rogue nodes information through $B_{msg}$ 
\State Terminate the rogue nodes detection algorithm
\end{algorithmic}
\end{algorithm}

 \begin{figure*}[htbp]
\centering
\includegraphics[width=500pt]{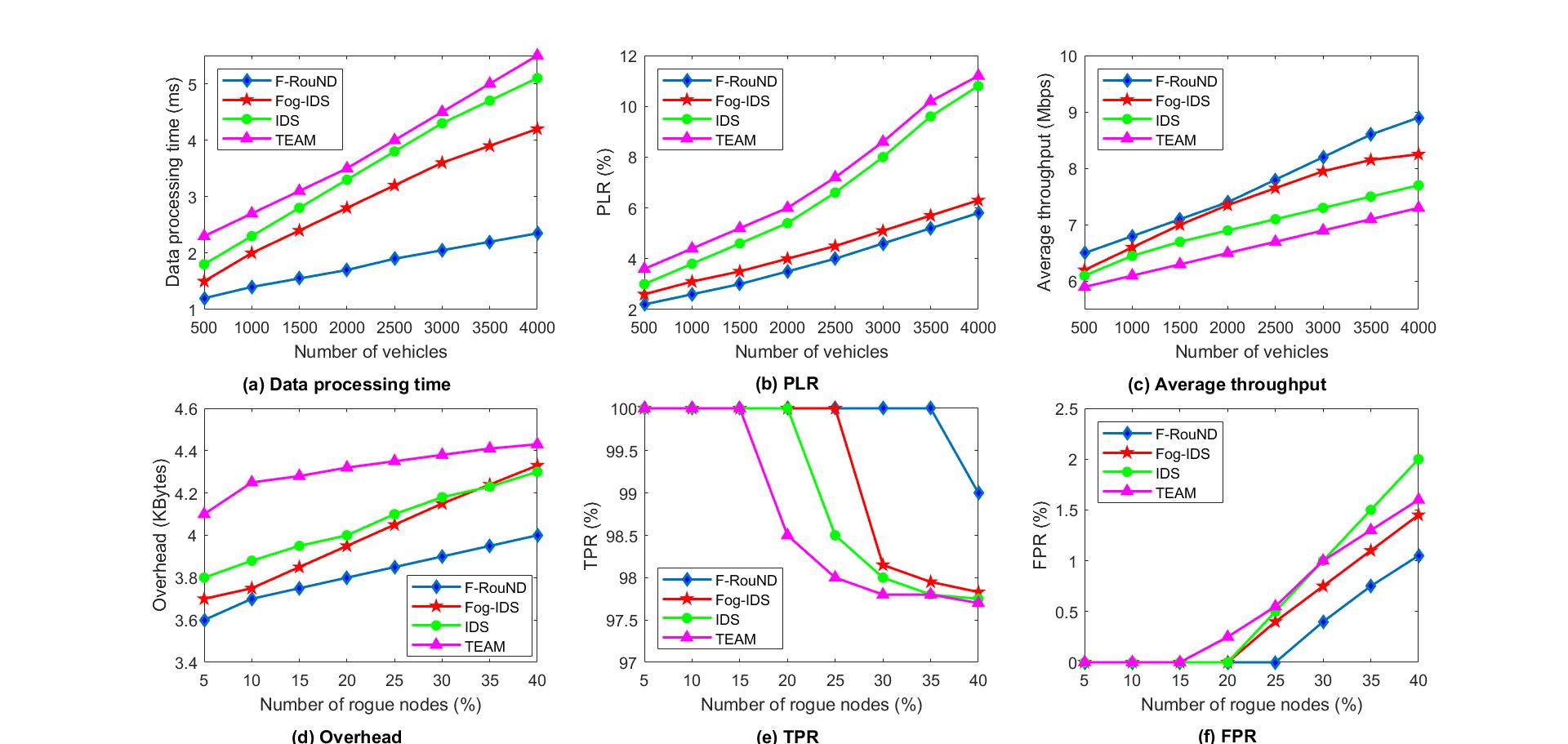}
\caption{Comparison of the F-RouND framework with Fog-IDS, IDS, and TEAM schemes: (a) data processing time, (b) PLR, (c) average throughput, (d) overhead, (e) TPR, (f) FPR. \label{fig3}}
\end{figure*}

\section {Performance Evaluation}
This section evaluates and analyzes the performance of our F-RouND framework discussed in Section 3. 
\begin{table}
\centering
\caption{Parameters used in Simulation of {F-RouND} Framework\label{tab1}}
\begin{tabular} [htbp] {|c|c|}

\hline

\textbf{Parameters} & \textbf{Values} \\    \hline \hline
Road length & 3 Miles \\    \hline
Number of vehicles & 500-4000 \\    \hline
Number of lanes & 2 \\    \hline
Vehicle speed &  30-65 Miles/hr\\    \hline
Beacon message size & 256 bytes \\    \hline
Transmission range & 500 m\\    \hline
Technique used & Fog computing \\    \hline
Protocol & IEEE802.11p \\    \hline
Simulator used & Omnet++, SUMO \\    \hline

\end{tabular}
\end{table}

\subsection {Analysis of F-RouND Framework}

In this analysis, we calculated the probability of failure. Failure of the system can occur due to loss of connectivity or a resource, insufficient capacity of fog, etc. The probability of system failure $P_{sysfail}$ is calculated by:

 \begin{equation} \label{eq13}
 \begin{split}
 P_{sysfail} = \sum_{i=0}^{N, t_{max}}{\left(N, t_{max} \atop i\right)}d_f^i(1-d_f)^{{N,t_{max}}^{-i}}
 \end{split}
 \end{equation} 
 
 Where $N$ is the number of vehicles, $t_{max}$ is the maximum time taken by the vehicles to get connected, and $d_f$ is the probability of success in the fog. Like quality of service (QoS), the probability of system failure contributes to the performance of the F-RouND framework. A minimum number of failures leads to the maximum performance of the fog.
 
\subsection {Simulation Setup}

The main objective of our simulation is to evaluate the performance of the F-RouND framework in the presence of rogue nodes (Section 3). We used OMNET++ and SUMO simulators to carry out the simulations. SUMO provides a trace of vehicular movements for a map imported from OpenStreetMap, while OMNET++ provides realistic modules, such as the packet loss model, node deployment model, etc. for realistic network behavior. We imported the city of Norman, Oklahoma, using OpenStreetMap into the SUMO simulator to generate vehicle traces. The output of the SUMO simulator, i.e., the trace of vehicles, is given as input to the OMNET++ simulator for rogue nodes detection. To assess the scalability and performance of the F-RouND framework, we performed a simulation with up to 4000 vehicles and 40\% rogue nodes. Table II summarizes the parameters used in the simulation. 

\subsection {Performance Metrics}

The simulations were performed based on the equations formulated in Section 3. We considered the following metrics to evaluate the performance of the F-RouND framework  and to compare our results with Fog-IDS, IDS, and TEAM schemes:

\begin{itemize}

\item Data processing time: The time needed by the guard node to compare and analyze vehicle speed in the beacon messages to detect rogue nodes in the region.

\item PLR: The ratio of the number of lost packets to the total number of packets sent across a communication channel.

\item Average throughput: Average rate of successfully broadcasted beacon messages across a communication channel. 

\item Overhead: The additional information exchanged between the vehicles to detect rogue nodes in the region.

\item True positive rate: The percentage of rogue nodes is accurately detected and classified as rogue nodes.

 \begin{equation} \label{eq13}
 \begin{split}
\text {TPR} = \frac{\text {No. of rogue nodes detected correctly}}{\text{Total no. of rogue nodes}}
 \end{split}
 \end{equation}

\item False positive rate: The percentage of honest nodes is incorrectly detected and classified as rogue nodes. 

 \begin{equation} \label{eq13}
 \begin{split}
\text {FPR} = \frac{\text{No. of honest nodes detected incorrectly}}{\text{Total no. of honest nodes}}
 \end{split}
 \end{equation}

\end{itemize}

\subsection {Simulation Results}

\textit{1) Data processing time:} The data processing time increases as the number of vehicles increases as more time needed to process the vehicle speed of all vehicles, as shown in Fig. 3a. In the F-RouND framework, the computation power of the guard node increases when the number of vehicles increases in the region as the OBUs of all vehicles are utilized in creating a fog results in 45\% lower processing time at high vehicle densities compared to [5-7] frameworks. In the 4000 vehicles simulation, the data processing time is 43\%, 52\%, and 57\% lower than Fog-IDS, IDS, and TEAM schemes, respectively. The evaluation of data processing time shows the F-RouND framework is scalable and can handle high vehicle densities.   

\textit{2) PLR:} The PLR is calculated against the number of vehicles and increases for all schemes when the number of vehicles increases from 500 to 4000. An increase in the number of vehicles increases the load on the network. When the network hits maximum capacity, packet drops occur. Also, PLR  increases due to the collision of some packets. In the F-RouND framework, the high computation power of the fog resulting in an optimum network capacity even with an increasing number of vehicles resulting in low PLR, i.e., 5\% PLR at high vehicle densities, as shown in Fig. 3b.

\textit{3) Average throughput:} The average throughput of the F-RouND framework is calculated against the number of vehicles, as shown in Fig. 3c. In the F-RouND framework, due to the high scalability of our dynamic fog (Section 3) and low PLR (Fig. 3b), the number of successfully broadcasted messages in the network increases with an increase in the number of vehicles resulting in high average throughput at all vehicle densities. In the 4000 vehicles simulation, the average throughput is 9\%, 17\%, and 23\% lower than Fog-IDS, IDS, and TEAM schemes, respectively.  

\textit{4) Overhead:} The overhead of the F-RouND framework increases with the increasing number of rogue nodes as an extensive hypothesis test needed to detect all rogue nodes in the region (Section 3C). Overhead is the additional information exchanged between the guard node and all other vehicles in the region to detect rogue nodes. In the F-RouND framework, vehicle speed in the beacon message is used to detect rogue nodes. Thus, the overhead of our F-RouND framework is 12\% lower compared to [5-7] schemes, even when the number of rogue nodes increased up to 40\%, as shown in Fig. 3d.

\textit{5) TPR:} The F-RouND framework identifies rogue nodes correctly (i.e., 100\%) up to 35\% rogue nodes in the region, as shown in Fig. 3e. As discussed in Section 3, the detection of rogue nodes is two-fold: first, the guard node compares the received speed values of all vehicles to detect rogue nodes in the region. Second, once the rogue nodes are identified, the guard node performs a hypothesis test to validate if the rogue nodes are correctly identified or not. However, when the number of rogue nodes is more than 35\%, the TPR decreases marginally to 99\%. It is difficult to detect the rogue node when the speed varies gradually. However, to generate either a false congestion scenario or catastrophic consequences, the target rogue node decreases the speed values suddenly. Thus, the F-RouND framework can detect rogue nodes even at high vehicle densities resulting in lower TPR compared to [5-7].  

\textit{6) FPR:} The increase in FPR deteriorates the performance of the proposed rogue node detection schemes. In the F-RouND framework, the rogue nodes detection relies only on the vehicle speed in beacon messages broadcasted by all vehicles in the region without using any trust scores or past vehicle data. Moreover, validation of the rogue nodes using hypothesis test results in 36\% lower FPR compared to existing schemes [5-7] even when the number of rogue nodes increases by up to 40\% in the region. For a network with 40\% rogue nodes, the FPR is 31\%, 51\%, and 38\% lower than Fog-IDS, IDS, and TEAM schemes, respectively, as shown in Fig. 3f.

\section{Conclusions and Future Work}

We studied challenges in rogue nodes detection, such as high processing delay, high network overhead, poor resource utilization, high FPR, and low TPR, notably when the number number of rogue nodes increases at high vehicle densities. To address these problems, we proposed an OBU-based fog computing technique, called F-RouND which ensures 45\% lower processing delay, 36\% lower FPR, and 12\% lower overhead at high vehicle densities compared to existing rogue nodes detection schemes [5-7]. We have analyzed the data processing time, PLR, average throughput, network overhead, TPR, and FPR, and performed a simulation using OMNET++ and SUMO simulators. Results showed that the F-RouND framework is efficient, scalable, and performs up to 38\% better than [5-7] schemes even when the number of rogue nodes increases by up to 40\% in the region. Moreover, the F-RouND framework does not depend on any roadside infrastructures like RSUs or trust scores or past vehicle data in rogue nodes detection, which is a major advantage compared to existing rogue nodes detection schemes. In the future, we plan to extend this work on vehicular social networks platforms. This can be done by simulating the environment of the social networks and then detecting the malicious information broadcasted using rogue nodes detection techniques.

\end{document}